\documentstyle[aps,floats,prc,epsfig]{revtex}
\newcommand{\beq}{\begin{equation}}
\newcommand{\eeq}{\end{equation}}
\newcommand{\bea}{\begin{eqnarray}}
\newcommand{\eea}{\end{eqnarray}}
\newcommand{\bef}{\begin{figure}}
\newcommand{\eef}{\end{figure}}
\newcommand{\bce}{\begin{center}}
\newcommand{\ece}{\end{center}}
\newcommand{\eg}{{\it e.g.}}
\newcommand{\ie}{{\it i.e.}}
\newcommand{\etal}{{\it et al.}}
\newcommand{\jj}{$J/\Psi$ }
\newcommand{\jjs}{$J/\Psi$'s }
\def\lsim{\mathrel{\rlap{\lower4pt\hbox{\hskip1pt$\sim$}}
    \raise1pt\hbox{$<$}}}         
\def\gsim{\mathrel{\rlap{\lower4pt\hbox{\hskip1pt$\sim$}}
    \raise1pt\hbox{$>$}}}         
\begin{document}
\twocolumn[\hsize\textwidth\columnwidth\hsize\csname
@twocolumnfalse\endcsname
\title{Thermal versus Direct $J/\Psi$ Production   
 in Ultrarelativistic Heavy-Ion Collisions}

\author{L. Grandchamp$^{a,b}$ and R. Rapp$^a$}

\address
{$^a$Department of Physics and Astronomy, State University of New York, 
    Stony Brook, NY 11794-3800, U.S.A. \\
$^b$IPN Lyon, IN2P3-CNRS et UCBL, 43 Bd. du 11 Novembre 1918, 69622 
Villeurbanne Cedex, France}

\date{\today} 

\maketitle
 \begin{abstract}
The production of $J/\Psi$ mesons in central collisions of heavy
nuclei is investigated as a function of collision energy. 
Two contributions are considered simultaneously: early (hard)
production coupled with subsequent suppression in a Quark-Gluon
Plasma, as well as thermal recombination of primordially
produced $c$ and $\bar c$ quarks at the hadronization transition.
Whereas the former still constitutes the major fraction of the 
observed $J/\Psi$ abundance at SpS energies, the latter dominates 
the yield at RHIC. The resulting excitation function for 
the number of $J/\Psi$'s over open charm pairs
exhibits nontrivial structure around $\sqrt s\simeq 30$~AGeV, 
evolving into a significant rise towards maximal RHIC energy. 
We study this feature within different (thermal) scenarios for 
$J/\Psi$ suppression, including parton-induced quasifree destruction 
as a novel mechanism.   
\end{abstract}
\vspace{0.3cm}
] 
\begin{narrowtext}
\newpage

A promising probe for  hot and dense QCD matter as created in
the early phases of   
heavy-ion reactions is the abundance of $J/\Psi$ mesons, 
measured via their decay branching into dilepton final states. 
Their production is expected to be significantly reduced in the case 
of Quark-Gluon Plasma (QGP) formation in sufficiently energetic 
collisions of large nuclei.
In such a picture, $J/\Psi$
mesons are exclusively formed primordially (\ie, upon first impact of the
colliding nucleons), and subsequently dissociated by (i) nuclear 
absorption, (ii) parton-induced destruction in
a QGP~\cite{Shu78} and/or Debye screening~\cite{MS86},
and (iii) inelastic scattering on ``comoving'' hadrons in the final 
hadron gas phase of the reaction.  
The identification of the plasma effect thus requires a reliable knowledge 
of both (i) and (iii). Nuclear absorption is appreciable and has been 
thoroughly investigated in $p$-$A$ and light-ion reactions, whereas the 
impact of hadronic interactions is not yet well under control, although 
its net effect seems to be rather moderate, see, \eg,  
Refs.~\cite{KS94,MBQ95,MM98,Vogt99,Ha00}. Within this framework the data of 
the NA50 collaboration at the SpS have been interpreted as evidence for QGP 
formation in the most central $Pb$(158~AGeV)-$Pb$
collisions~\cite{na50psi}.   

Recently, an alternative view of $J/\Psi$ production in heavy-ion 
reactions has been put forward. Prompted by the observation that the 
$J/\Psi$ yield per charged hadron is remarkably constant as a function
of impact parameter, it has been argued in Ref.~\cite{GG99} that all  
$J/\Psi$'s are created statistically at the hadronization transition. 
The deduced temperature of $T\simeq 175$~MeV is well in line with the 
so-called chemical freeze-out of {\em light} hadron 
production~\cite{pbm96-99,YG99,Beca}. However, no reference is made to 
an underlying mechanism for $c\bar c$ creation. In a somewhat different 
approach, Braun-Munzinger and Stachel have extended their thermal 
model analysis~\cite{pbm96-99} to include (open and hidden) charm 
hadrons~\cite{BS00-1} (see Ref.~\cite{BS00-2} for an update of this 
analysis). Together with the (dynamically well justified)
proposition that $c\bar c$ pairs at SpS energies are exclusively 
produced primordially, the $J/\Psi$ abundance in sufficiently central 
$Pb$-$Pb$ collisions  can be accounted for by  
statistical recombination of $c$ and $\bar c$ quarks at the 
earlier inferred hadro-chemical freeze-out without introduction of 
new parameters. 
In a subsequent analysis~\cite{GKSG00}, this approach was reiterated
using a more complete set of charmed hadrons and enforcing exact (local)
charm conservation within a canonical-ensemble treatment. Requiring 
to reproduce the NA50 measurements for $J/\Psi$ production, an open-charm
enhancement factor of $\sim$~3 relative to $N_{p}^{4/3}$ times the value 
in $N$-$N$ collisions was deduced for central $Pb$-$Pb$ collisions 
($N_{p}$: number of participant nucleons). This coincides 
with the enhancement needed to explain the NA50 intermediate-mass region 
(IMR) dilepton spectra solely in terms of increased open-charm 
production\footnote{In this paper we adopt a scenario without  
any ``anomalous'' open-charm enhancement, attributing  
the NA50 IMR dilepton excess to thermal radiation~\cite{RS00,GKP00}.}. 
As first pointed out in Ref.~\cite{BS00-1}, the application of the
thermal production framework to RHIC energies could 
in fact lead to an enhancement of $J/\Psi$'s over its primordial
production rate (see also Ref.~\cite{TSR00}). 

One of the main assumptions in the thermal model analyses is that 
primordial production of $J/\Psi$'s is absent, \ie, they either 
do not form or are fully suppressed before the hadronization transition. 
Under SpS conditions with supposedly rather short plasma lifetimes of 
$1$-$2$~fm/c, this assertion is, however, not easily realized.   
In this article we therefore attempt a combined description 
of thermal and primordial production, the latter being subjected 
to nuclear absorption and plasma dissociation in an
expanding fireball model around midrapidities. The focus will be 
on central $Pb$-$Pb$ ($Au$-$Au$) collisions at varying $CMS$ energy,
covering the SpS and RHIC regime. We fix the participant number at 
$N_{p}\simeq360$ (corresponding to an impact parameter $b\simeq 1.5$~fm)
to avoid complications associated with transverse-energy fluctuations 
in the most central collisions~\cite{CFK00,BDO00,HKP00}. 
Also, $c\bar c$ pairs will be allowed to coalesce into 
charmonium states only over a limited range of rapidities.

Let us start by discussing our implementation of the contribution
from primordial production and subsequent suppression. As in 
Ref.~\cite{BS00-1}, we assume that the production of $c\bar c$ pairs
entirely occurs in primary (hard) nucleon-nucleon collisions 
(secondary creation in a partonic medium has been shown to be negligible 
even at full RHIC energy ~\cite{Re00}).
In free space a certain fraction $F_{\Psi}$ of these pairs combines into 
a final number $N_{J/\Psi}^{dir}$ of directly produced $J/\Psi$ mesons 
($F_{\Psi}\simeq2.5$\% in the SpS energy regime~\cite{Gav95}).
In a heavy-ion environment, the first phase of suppression is 
characterized by interactions of the (pre-resonant) bound state with 
interpenetrating nucleons. It leads to a  rather well understood 
$(N_{p1} N_{p2})^\alpha$ suppression of the cross section with 
$\alpha=0.92\pm0.01$ as inferred from $p$-$A$ 
and $A$-$B$ reactions with light projectile nuclei (see also
Ref.~\cite{HHK00}).
This factor equally applies to other charmonium states 
($\chi_c$, $\Psi'$), which contribute  
via their decay branchings into $J/\Psi+X$ final states (``feeddown'').
We model this so-called nuclear absorption using a Glauber model with 
a phenomenological constant cross section 
$\sigma_{\Psi N}\simeq5.8$~mb~\cite{NA50QM01}.

In the second phase of suppression -- the QGP -- charmonium destruction
has been discussed in both static screening-type pictures as well as
dynamical ones via inelastic collisions with partons, most notably the QCD 
analogue of photo-dissociation,  $g+J/\Psi \to c \bar c$~\cite{Shu78,BP79}.  
Within an expanding fireball model (see below) we follow the dynamical
picture, accounting, however, for a reduced $J/\Psi$ binding energy. 
The dissociation rate is calculated from 
\beq
\Gamma_{diss}=\sum\limits_{i=q,g} \;
\int\limits_{k_{min}}^{\infty} \frac{d^3k}{(2\pi)^3} \
f^{i}(k;T) \ \sigma_{diss}(s)
\eeq
with $k_{min}$ denoting the minimal on-shell momentum of a quark or gluon
from the heat bath necessary to dissolve an in-medium charmonium bound
state into a (free) $c\bar c$ pair. The binding is characterized by a
temperature-dependent dissociation energy $E_{diss}$ taken from
Ref.~\cite{KMS88} using a Debye screening mass $m_D^2=g^2 T^2$ with a 
typical $g\simeq 1.7$; this entails $E_{diss}(T=180{\rm MeV})\simeq 220$~MeV,
dropping to $\simeq$~100~MeV at $T=240$~MeV but crossing zero only around 
$T_{Debye}\simeq$~400~MeV. With such a decrease in the $J/\Psi$
binding energy, the break-up 
kinematics render the photo-dissociation process increasingly 
inefficient. For a loosely bound charmonium state, 
a more important mechanism turns out to be given by inelastic parton 
scattering, $g(q,\bar q)+ J/\Psi \to g(q,\bar q) + c + \bar c$.   
We evaluate the respective cross sections in quasifree approximation 
using leading-order QCD~\cite{Com79} for $gc\to gc$ ($qc\to qc$) and 
the appropriate break-up kinematics.  In addition to a gluon
screening mass, thermal quasiparticle masses for light quarks 
($m_{u,d}^2=g^2T^2/6$, $m_s^2=m_0^2 +g^2T^2/6$) and gluons 
($m_g^2=g^2T^2/2$)~\cite{Bel96} are included.
The resulting dissociation times, $\tau_{diss}=\Gamma_{diss}^{-1}$,
are shown in Fig.~\ref{fig_tau} and compared to calculations with a 
constant break-up cross section of 1.5~mb, as well as to the 
photo-dissociation mechanism {\em without} medium effects
in the $J/\Psi$ bound state energy (as it has been employed in the
literature before).
\begin{figure}[!t]
\vspace{-0.2cm}
\begin{center}
\epsfig{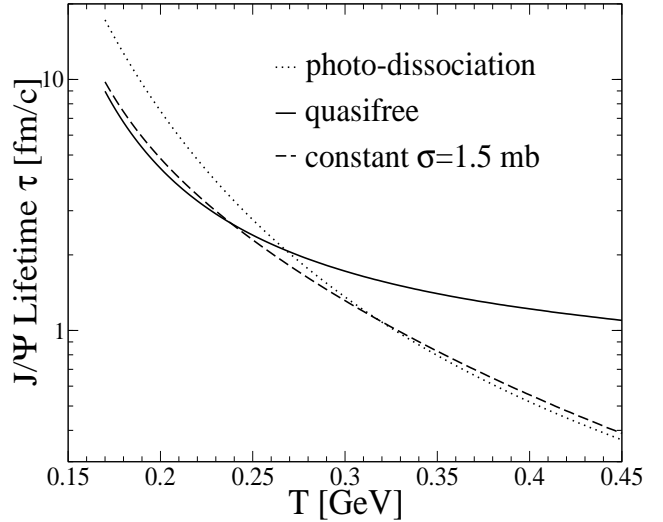} 
\vspace{0.2cm} 
\caption{Dissociation times for a $J/\Psi$ in a QGP as a
  function of temperature. The full curve corresponds to the 
  leading-order QCD process for quasifree $g,q+c\rightarrow g,q+c$
  scattering with {\em in-medium} $J/\Psi$ bound state energy. 
  The dashed curve represents a constant cross-section
  $\sigma_{diss}=1.5$~mb and the dotted curve results from  
  photo-dissociation, $gJ/\Psi \rightarrow c\bar{c}$, assuming
  the {\em vacuum} dissociation energy.} 
\label{fig_tau}
\end{center}
\vspace{-0.3cm}
\end{figure}
At temperatures relevant for SpS conditions, the quasifree
dissociation process (full line) is more efficient than
photo-dissociation (dotted line). At higher temperatures, it becomes
less efficient due to an increasing gluon screening mass which suppresses the
$t$-channel exchange graphs for $g(q,\bar q)+ J/\Psi \to g(q,\bar q) + c + \bar c$. 

The temperature (time) dependence of the dissociation rate has to be 
coupled with a model for the space-time evolution of the reaction 
dynamics.  To facilitate the calculations we here employ a thermal 
fireball description in line with earlier analyses of  
dilepton radiation at both SpS~\cite{RS00,RW99} and
RHIC~\cite{Ra01}. Let us briefly recall its essential elements.    
After initial impact of two colliding $Au$ (or $Pb$)
nuclei the system is assumed to be thermalized after a formation 
time $\tau_0$. Thereafter, the fireball undergoes isentropic 
expansion characterized by conserved entropy and (net) baryon number
which defines a thermodynamic trajectory in the $\mu_B-T$ plane of the 
phase diagram.
Above the critical temperature $T_c$ a (quasiparticle) QGP equation 
of state is used, and a resonance hadron gas one below. The transition 
is modeled by a standard mixed phase construction~\cite{McLe86}, 
\beq
S/V(t)=fs_{HG}(T_c) + (1-f)s_{QGP}(T_c) \ ,
\eeq
justified for a sufficiently sharp increase of the entropy density 
around $T_c$ ($f$: fraction of matter in the hadronic phase, 
$s_{HG}$ ($s_{QGP}$): entropy density in the hadronic (plasma) phase). 
$S$ denotes the total entropy in the considered rapidity interval, 
and $V(t)$ the time dependent volume therein, which we simulate 
by two fireballs with cylindrical expansion as
\beq
V(t)=2\left(z_0+v_z t+a_z \frac{t^2}{2}\right)
\pi\left(r_{\perp}+a_{\perp}\frac{t^2}{2}\right)^2 \ .  
\eeq
The parameters $\{v_z, a_z, a_{\perp}\}$ are adjusted 
to finally observed flow velocities in connection with total 
fireball lifetimes of around 15~fm/c. 
The $\sqrt{s}$-dependence of the collisions is constructed as follows:
the formation time $\tau_0$ is taken to be $\sim 1$~fm/c at SpS and 
$\sim\frac{1}{3}$~fm/c at 
full RHIC energy (with a powerlike interpolation in $\sqrt{s}$), resulting
in initial temperatures of $T_0\simeq205$~MeV and 390~MeV, respectively.  
We assume the transition temperature $T_c$ to smoothly increase from 
$T_c=170$~MeV at $\sqrt{s}=17.3$~GeV to $T_c=180$~MeV at
$\sqrt{s}=200$~GeV (with approximately constant critical
entropy density in the hadronic phase). 
SpS and first RHIC data (at $\sqrt{s}=56$~GeV and $130$~GeV) on total
multiplicities~\cite{phobos00} and $\bar{p}/p$, $\bar{\Lambda}/\Lambda$ 
ratios~\cite{star00} are used to estimate the total entropy as well as
baryon and strange-quark chemical potentials.  
This description of the collision gives results
consistent with hydrodynamical calculations, \eg,  a pure QGP
lifetime of $\sim 1.5$~fm/c (3.5~fm/c) at SpS (RHIC),
and a mixed phase until  $\sim 5$~fm/c ($7.5$~fm/c).
The QGP suppression factor ${\cal S}_{QGP}$ of $J/\Psi$ mesons follows 
from integrating the dissociation rate over the space-time evolution.  
The resulting direct yields per central collision (including nuclear
absorption) are listed in Tab.~\ref{tab:thermal}.

The second source of charmonium states originates from thermal 
production at hadronization\footnote{In principle, $J/\Psi$ formation 
can also occur above $T_c$~\cite{TSR00} through the reverse of the 
dissociation process, \ie, $c\bar c g \to J/\Psi g$; however, due to 
the smallness of the $J/\Psi$ binding energy in the plasma, implying 
large formation times, we neglect formation above $T_c$, thus possibly 
underestimating thermal production somewhat.}. The underlying 
picture~\cite{BS00-1} is a statistical coalescence of $c$ and $\bar{c}$ 
quarks at $T_c$. In thermal models, hadron production is 
determined  by the available phase space at $T_c$. 
The total number of particle species $j$ then is
\begin{equation}
  \label{eq:part_num}
  N_j  =  \frac{d_jV}{2\pi^2}\int\limits_0^{\infty} p^2
  dp\left[\exp\left(\frac{\sqrt{p^2+m_j^2}-\mu_j}{T}\right)\pm1\right]^{-1} \ , 
\end{equation}
where $d_j$ denotes the degeneracy factor, $\mu_j$ the pertinent
chemical potential,  $\mu_j = B_j\mu_B + s_j\mu_s + c_j\mu_c$, and $V$
the hadronic fireball volume at $T_c$. Since at CERN-SpS
$N_{c\bar{c}} \ll 1$, exact charm conservation is enforced 
within a canonical-ensemble treatment (see, \eg, Refs.~\cite{Shu75,RT80}). 
We include all known charmed hadrons~\cite{pdg00} and fix the 
number $N_{c\bar{c}}^{dir}$ of $c\bar{c}$ pairs from primordial 
$NN$ collisions in our restricted rapidity range as given by  
PYTHIA computations~\cite{pythia} upscaled by an empirical $K$ factor, 
$K \simeq 5$~\cite{AD00}. This necessitates the introduction
of a fugacity $\gamma_c=\gamma_{\bar{c}}$ for charm and
anticharm quarks according to  
\begin{equation}
  \label{eq:gamma}
  N_{c\bar{c}}^{dir} = \frac{1}{2}\gamma_c
  N_{open}\frac{I_1\left(\gamma_c N_{open}\right)}{I_0\left(\gamma_c
      N_{open}\right)} + \gamma_c^2 N_{hidden},
\end{equation}
where $N_{open}$ ($N_{hidden}$) denotes the thermal abundance  
of open (hidden) charm hadrons ($I_{0,1}$ are modified Bessel functions). 
Hence, the total contribution to statistical $J/\Psi$ production
(including strong and electromagnetic feeddown) follows as
$\langle J/\Psi \rangle = \gamma_c^2 N_{J/\Psi}^{tot}$, 
cf.~Tab.~\ref{tab:thermal} for selected collision energies.
\begin{table}[!h]
\vspace{-0.1cm}
\begin{tabular}{lcccc}
$\sqrt{s}$ [GeV] & 17.3 & 56 & 130 & 200 \\
\hline 
${\cal S}_{QGP}$ & 0.66 & 0.50 & 0.33 & 0.23 \\
$N_{J/\Psi}^{dir}$ [$10^{-3}$] & 0.45 & 2.41 & 3.68 & 3.25 \\
\hline
$N_{c\bar{c}}^{dir}$  & 0.17 & 2.34 & 7.53 & 10.86 \\
$\gamma_c$ & 0.82 & 2.46 & 4.85 & 5.62 \\
$\langle J/\Psi \rangle$  [$10^{-3}$] & 0.16 & 4.31 & 27.30 & 45.22 \\
\end{tabular}
\caption{Direct and statistical production of \jjs per central
  collision ($N_p=360$) at various $cm$ energies in a fixed rapidity
  window covered by two fireballs. $N_{J/\Psi}^{dir}$: number of
  primordial $J/\Psi$'s remaining after nuclear absorption and plasma
  suppression (${\cal S}_{QGP}$). $\gamma_c$: charm quark fugacity
  deduced from (\protect\ref{eq:gamma}) based on $N_{c\bar{c}}^{dir}$
  primordial $c\bar{c}$ pairs. $\langle J/\Psi \rangle$: number of
  statistically produced \jjs.}
\label{tab:thermal}
\vspace{-0.1cm}
\end{table}

Combining the two sources of \jjs (the suppressed direct as well as   
the statistical (or thermal) production), we calculate 
excitation functions from SpS to RHIC energies. 
Fig.~\ref{fig:enhance} displays the ratio of the observed number
of \jjs over the primordially produced one. 
\begin{figure}[!b]
\vspace{-0.5cm}
\begin{center}
\epsfig{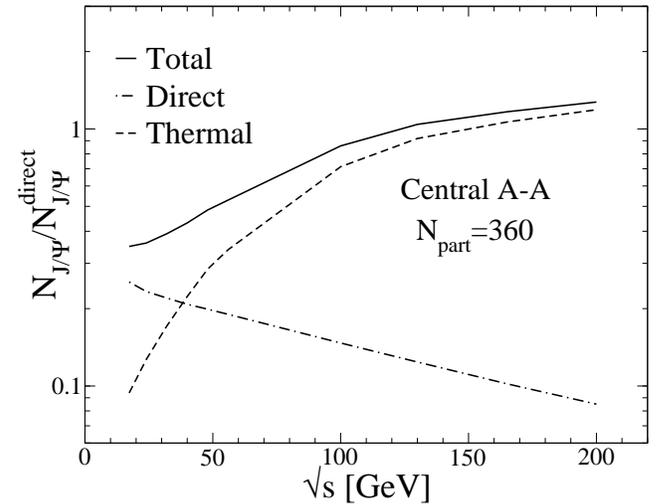} 
\vspace{0.3cm}
\caption{Ratio of the number of observed \jjs over the number of
  primordially produced ones (full curve) in a two-fireball model
  around midrapidity. The dashed (dash-dotted)
  curve corresponds to the \jj yield from
  statistical coalescence at hadronization (direct
  production with nuclear absorption and QGP suppression).}
\label{fig:enhance}
\end{center}
\vspace{-0.3cm}
\end{figure}
The thermal source accounts for one fourth of the yield at SpS, but 
dominates at higher energies implying a possible \jj enhancement 
at RHIC~\cite{BS00-1,TSR00}.

In anticipation of open-charm measurements at both 
RHIC and SpS~\cite{na60}, one can make closer contact to 
observables by plotting the ratio of the final 
number of \jjs over the primordial number of $c\bar{c}$
pairs, cf.~Fig. \ref{fig:ccplot}.   
\begin{figure}[!h]
\vspace{-0.1cm}
\begin{center}
\epsfig{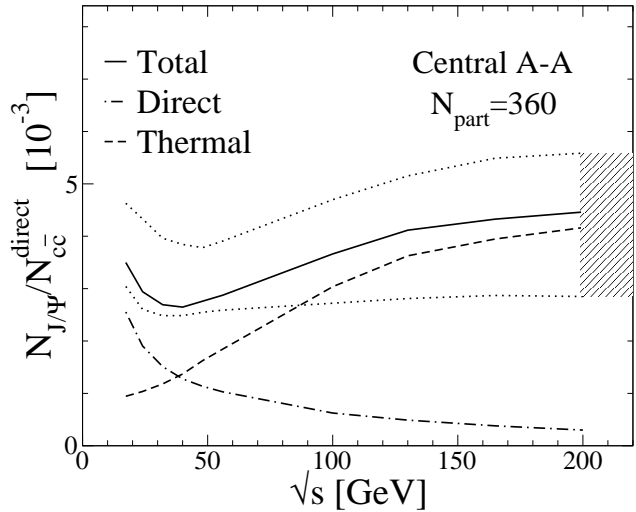} 
\vspace{0.2cm}
\caption{Ratio of the number of observed \jjs over the number of
  primordial $c\bar{c}$ pairs (full curve) in a two-fireball model
  around midrapidity. The dashed (dash-dotted) curve shows the statistical 
  contribution (direct production with nuclear and QGP suppression) to
  this ratio. The band enclosed by the dotted lines reflects the uncertainty 
  on the initial charm and \jj production as explained in the text.}
\label{fig:ccplot}
\end{center}
\vspace{-0.3cm}
\end{figure}
This ratio exhibits a minimum for $\sqrt{s}\simeq 30$~GeV when the thermal 
and direct contributions are about equal. 
A similar minimum stucture has been found in Ref.~\cite{Go00}, where, however, the 
$J/\Psi$ yield at all energies was entirely attributed to statistical 
production.

The largest sensitivity in our calculation is attached to primordial 
$c\bar{c}$ production as indicated by the band enclosed by the dotted lines
in Fig.~\ref{fig:ccplot}. 
The lower limit is estimated from next-to-leading order pQCD
calculations~\cite{McGAU95} in connection with a lower bound 
in  \jj production from available data  (supplemented by a
phenomenological fit at higher energies~\cite{L96}). The upper
limit is obtained from
a PYTHIA calculation for $c\bar{c}$ production using GRV-HO structure 
functions (which tends to give the largest yield towards RHIC energies)
in connection with \jj production from Ref.~\cite{Gav95}.
We also checked that there is only moderate sensitivity to 
variations in the hadronization temperature: decreasing $T_c$ to 170~MeV
at $\sqrt{s}=~200$~GeV (with an accompanying increase in the hadronization 
volume, but at fixed $N_{c\bar c}$) entails a 10\% larger yield of thermal 
$J/\Psi$'s (the decrease in the thermal density is overcompensated by 
the increase in volume and, more importantly, by the higher charm-quark
fugacities). 
Within the uncertainties the plotted ratio persists to exhibit 
a very different behavior (\ie, an increase with $CMS$ energy) 
from the one expected in the standard scenario of \jj {\em suppression}.

Finally, we investigate the sensitivity of the minimum structure with
respect to different QGP suppression mechanisms. Upon replacing the 
quasifree destruction process by the gluon photo-dissociation process 
shown in Fig.~\ref{fig_tau} (dotted line), we observe a slight overall 
increase in the yield without significant alteration of the shape. 
Thirdly, in a more extreme scenario based on Debye-screening, we assume 
$J/\Psi$ mesons to be entirely suppressed if they are formed in a 
region with initial energy density $\epsilon_0(r)>\epsilon_{Debye}$ 
(along the lines of Ref.~\cite{BO96}). Within the Glauber
model, the spatial distribution of primordial $J/\Psi$'s is inferred 
from the nuclear thickness function $T_{AB}(r)$ (characterizing 
the number of $N$-$N$ collisions), whereas the energy-density profile 
is taken to be proportional to the density of participants in the 
transverse plane. We fix $\epsilon_{Debye}$ to obtain a suppression 
consistent with the NA50 data at $\sqrt{s}=17.3$~GeV (translating 
into $T_{Debye}\simeq 220$~MeV). As expected,
the pertinent excitation function exhibits a stronger suppression
pattern with increasing $\sqrt{s}$, generating a more pronounced 
minimum structure (at similar position) in the ratios 
$N_{J/\Psi}/N_{J/\Psi}^{dir}$ and $N_{J/\Psi}/N_{c\bar c}^{dir}$ 
than found with dynamical  dissociation processes, 
cf.~Fig.~\ref{fig:debye}.  
\vspace{0.9cm}
\begin{figure}[h]
\vspace{0.0cm}
\begin{center}
\epsfig{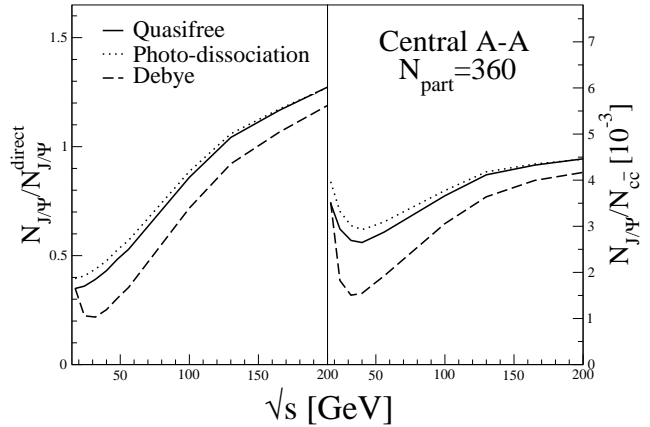}
\vspace{0.2cm}
\caption{Comparison of Debye-screening (dashed curves), photo-dissociation 
(dotted curves) and quasifree dissociation (full curves) of $J/\Psi$'s in 
a thermalized QGP at various collision energies, combined with thermal 
production. Photo-dissociation (with vacuum $J/\Psi$ binding energy) gives 
results very similar to quasifree suppression (with in-medium $E_{diss}$)
whereas Debye screening induces stronger suppression  
reflected in a more pronounced minimum around $\sqrt{s}=30$~GeV.}
\label{fig:debye}
\end{center}
\vspace{-0.3cm}
\end{figure}

In summary, we have proposed a ``combined'' approach to 
evaluate \jj yields in heavy-ion collisions which includes 
(i) a direct contribution of prompt \jjs subject to nuclear and
Quark-Gluon Plasma absorption and, (ii) a thermal
contribution of \jjs emerging from recombination of $c$ and $\bar{c}$ 
quarks at hadronization. The employed framework is consistent with 
earlier calculations of thermal dilepton spectra; in particular, no
enhancement in open-charm production has been invoked.
The resulting \jj excitation function exhibits a transition from mostly 
primordial to dominantly thermal production when going from SpS to
RHIC. Such an interplay could be mapped out by a systematic 
variation in collision energies accessible at RHIC.  
The predicted increase in \jj yields will render \jj {\em suppression} 
difficult to identify as a QGP signature. However, 
the excitation function might serve as a sensitive probe of the
hadronization dynamics at the QCD phase transition, provided an accurate
knowledge of primordial $c\bar{c}$ abundances. 
We also note that our description might imply significant changes in the \jj
transverse-momentum distributions. At the highest RHIC energies, one
expects essentially thermal shapes (accompanied by a flow component 
from the QGP phase), which should be distinguishable  
from hard production prevalent at the SpS.

In this work, we did not address phenomena in the
later hadronic  stages of the collision, \eg, possible consequences  
of (the approach towards) chiral symmetry restoration via in-medium
modifications of $D$-meson masses which might play an important role 
for the $\Psi'/\Psi$ ratio~\cite{SSZ97,STST}. 
An extension of our approach along these lines 
together with a detailed comparison to available data 
on centrality and projectile dependence is in progress~\cite{GR01}. 

\acknowledgments
We thank E.V. Shuryak for valuable comments throughout the course
of this work and J. H\"ufner for an interesting discussion. 
This work was supported by the U.S. Department of Energy 
under Grant No. DE-FG02-88ER40388.

\end{narrowtext}
\end{document}